# Infrarotspektroskopie plausibel

Petra Schulz

Theodor-Francke-Weg 65, D-38116 Braunschweig, Deutschland

**Abstract**

The rotation vibration spectra of small molecules can be described mathematically completely. Unfortunately, no vivid interpretation of the observed transitions exists. A qualitative interpretation attempt is undertaken for symmetric top molecules as example. For the first time the rotation changes in the case of parallel and vertical bands are explained mechanistically. A plausible photon model is represented simultaneously.

**Kurzfassung**

Die Rotations-Schwingungs-Spektren kleiner Moleküle lassen sich mathematisch vollständig beschreiben. Was fehlt, ist eine anschauliche Deutung der beobachteten Übergänge. Es wird ein qualitativer Deutungsversuch unternommen am Beispiel von symmetrischen Kreiselmolekülen. Parallel- und Senkrecht-Schwingungen mit den zugehörigen Rotationsübergängen in und quer zur Figurenachse samt einem Photonenmodell werden erstmals plausibel dargestellt.

## 1. Einleitung

Bei der Infrarotspektroskopie (Rotations-Schwingungs-Spektroskopie) schlucken Moleküle Photonen im Energiebereich der Wärmeenergie. Für eine bestimmte Molekülsorte ist dabei ein linienreiches Spektrum zu beobachten. Die quantitative Behandlung von spektroskopischen Daten ist dank mathematischer Modelle auf der Basis der Quantenmechanik bis in alle Einzelheiten bekannt. Die qualitative Deutung bleibt jedoch rätselhaft, obwohl wir im Zeitalter der Nanotechnologie leben. In der vorliegenden Arbeit wird ein qualitativer Deutungsversuch unternommen. Dieser Erklärungsversuch ist z. B. wichtig, wenn es gilt, Streuung an einzelnen Molekülen oder auch an vielen Teilchen in einem Gitterverband unter kohärentem Laserbeschuß zu verstehen.

## 2. Was sind Photonen?

### 2.1 Das Photon des freien Elektrons

Große Frage nun: Was sind denn eigentlich Photonen? Bei der DPG-Tagung in Dresden habe ich versucht, Masse, Ladung und Spin plausibel zu definieren und benutzte dabei ein anschauliches Photonenmodell [7]. Angenommen, es sollen Elektronen mit Photonen beschossen werden. Dann sind die einfachsten Photonen (etwa als Stoßpartner von ruhenden, bahndrehimpulslosen Elektronen) am ehesten vergleichbar mit Geschosskugeln aus einem Gewehr. Photonen haben somit

a) einen Vorwärtsdrang, einen Bahnimpuls, (in Abb. 1 dargestellt als waagerechter Pfeil). Photonen sind allerdings schneller als Gewehrkugeln, denn sie fliegen mit Lichtgeschwindigkeit. Auf den - wie auch immer gearteten Bahnimpuls - will ich im folgenden aber nicht eingehen, dafür aber auf den Spin der Lichtteilchen.

b) Photonen sind gedrallt wie heimtückische Fußbälle oder Billardkugeln, also sie rotieren um die eigene Achse, sie besitzen einen Spin, (in Abb. 1 dargestellt als gebogener Pfeil). Der Spin befindet sich im einfachsten Fall entweder in der bzw. entgegengesetzt zur Ausbreitungsrichtung – also ein longitudinal polarisiertes Photon (s. Abb. 1a) - oder senkrecht zur Ausbreitungsrichtung – also ein transversal polarisiertes Photon (s. Abb. 1b).

In der Maxwell-Theorie werden nur die transversal polarisierten Photonen berücksichtigt. Das ist gewiss eine Unzulänglichkeit, zumindest in der (Molekül)-Spektroskopie. Natürlich kommen auch Linearkombinationen der beiden Polarisationszustände im beliebigen Verhältnis gleichzeitig vor.

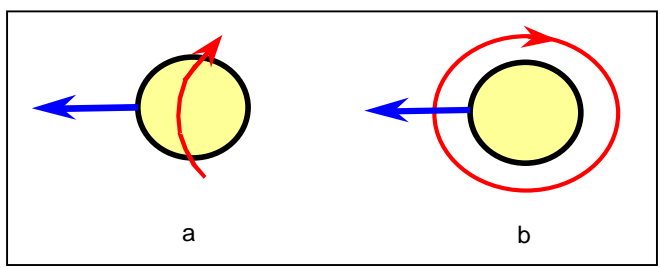

**Abb. 1**: Photon mit Vorwärtsimpuls und Spin
a) Beispiel eines longitudinal polarisierten Photons
b) Beispiel eines transversal polarisierten Photons

## 2.2 Die Photonen der Moleküle

Die eben vorgestellte anschauliche Vorstellung über Photonen soll im Prinzip auch beibehalten werden, wenn die Moleküle durch ein Photon angeregt werden wie beispielsweise bei der Infrarotspektroskopie. Die Änderungen im Molekül möchte ich qualitativ anschaulich sehen und nicht abstrakt durch die Wellenfunktion. In der schriftlichen Fassung meines letzten DPG-Vortrags in Bremen [8] habe ich die Photonen ganz grob als Negativ von den Teilchen definiert. (Das soll nur eine Zusatzerklärung dafür sein, weshalb man die Gruppentheorie in der Spektroskopie benötigt.) Vielleicht sollte ich das noch etwas präzisieren: Die absorbierten oder auch die später re-emittierten Photonen müssen genauso strukturiert sein wie die angeregten oder das abgeregten Teile in den Partikeln. Somit brauche ich mich bei den Stoßprozessen nicht um die Strukturfunktion (S-Matrix) der Teilchen zu kümmern.

Ein Infrarotspektrometer schießt in und quer zur Ausbreitungsrichtung der Photonen mit links-oder/und rechtsdrehenden Photonen auf die Moleküle. Ein Photon kann seine Rotationsenergie loswerden, wenn es ein Molekül anschießt. Alle Moleküle sind nämlich potentielle Kreisel.

## 2.3 Durchlässigkeit der Teilchen für Photonen

Photonen können bei Stoßprozessen durch Teilchen hindurchgehen. Ich erinnere dabei an das Toc-Toc-Spielzeug, die Pendelkette, bzw. auch noch an den Stafettenlauf der Photonen durch die Elektronen bei der elektrischen Leitung in Metallen. Wenn man vom ersten und letzten Teilchen absieht, werden nur Photonen bewegt. Photonen sind für viele Teilchen durchlässig. Diese Transparenz der Teilchen für Photonen ist dabei vor allem vom jeweiligen Energiezustand der Teilchen abhängig.

## 2.4 Die Wellenfunktion des Photons

Ich weiß: die Wellenfunktion des Einzelteilchens ist genau wie die Brechzahl [8] eine reelle Größe. Und ich extrapoliere mit gutem Recht: auch die Wellenfunktion eines Ensembles von vielen Teilchen ist bei der Wechselwirkung mit Photonen garantiert reell.

## 3. Rotationsspektroskopie

### 3.1 Zustandekommen eines Rotationsspektrums

An zweiatomigen (stäbchenförmigen) Molekülen wird keine Rotation um die Figurenachse durch spektroskopische Methoden beobachtet, sondern (im geeigneten Fall bei dipolmoment-behafteten Molekülen) nur um die Senkrechte der Bindungsachse der beiden Atome. Deshalb wollen wir vor vornherein zu mehratomigen Molekülen übergehen. In Abb. 2 ist ein Molekülkreisel mit dreizähliger Symmetrieachse gezeichnet. Das kann z. B. Methylchlorid ($CH_3Cl$) sein oder vielleicht auch Silylchlorid ($H_3SiCl$). Solch ein Molekül gleicht einem Kinderkreisel, der auf dem Kopf, seiner Spitze, steht. Angenommen, es fliegen zwei Photonen auf das Molekül zu: Photon p (p für parallel) und Photon s (s für senkrecht). Wegen der Übersichtlichkeit wurde in Abb. 2 die Struktur des Photons unterdrückt. Wenn solch ein Kreisel bäuchlings vom Photon s getroffen wird, dann wird er senkrecht zu seiner Figurenachse kreiseln, deshalb taufe ich diese Bewegung als **Senkrechtrotation**. Wenn Photon p sich köpflings nähert, wird es den Kreisel um seine Symmetrieachse rotieren lassen, diese Bewegung kennen wir vom Kinderkreisel her. Da die Rotationsachse (Figurenachse) parallel (sogar direkt in der Figurenachse liegt), taufe ich die erfolgende Bewegung **Parallelrotation**, s. Abb. 2. Senkrecht- und Parallelrotation sind übrigens keine gängigen Bezeichnungen. Prinzipiell gibt es übrigens zwei Senkrecht-Rotationen, die eingezeichnete und die dazu senkrechte. Im Falle symmetrischer Kreisel sind die beiden Senkrecht-Rotationen entartet (energetisch gleichwertig).

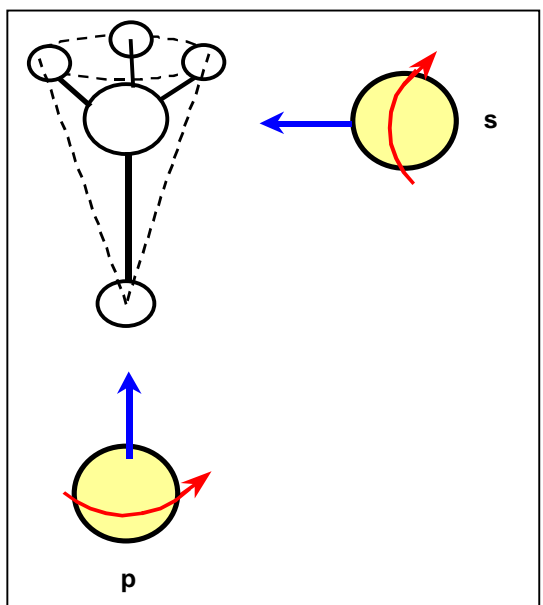

**Abb. 2:** Der Kreisel $CH_3Cl$ oder $H_3SiCl$ wird köpflings – parallel - zur Figurenachse durch das Photon p wie bei der Rotation eines Kinderkreisels oder/(und) bäuchlings - senkrecht dazu - von dem Photon s getroffen.

Folgende Vermutung erscheint berechtigt: Die erwähnten Vorgänge klappen nicht gleich auf Anhieb, denn sie benötigen eine gewisse Anlaufzeit. Molekül und Photon müssen diese regelmäßig koordinierten Bewegungen nämlich erst üben. Dafür sind mehrere Ansätze oder Nachbesserungen nötig.

Abb. 3a zeigt vereinfacht eine Parallelrotation und Abb. 4a eine Senkrechtrotation. Zur Vereinfachung wurden die Kreiselmoleküle nur noch in der Seitenansicht als Dreiecke dargestellt. In Abb. 3b und 4b wird nach weiterer Vereinfachung nur die Projektion des Rotationsumlaufs von der Seite eingezeichnet.

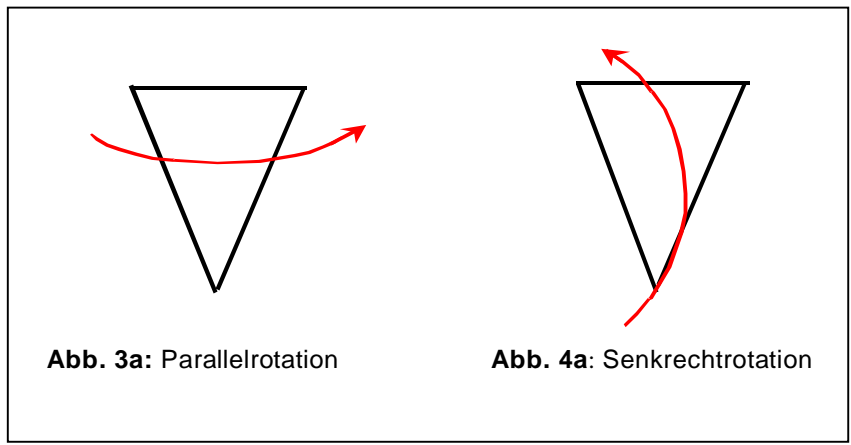

**Abb. 3a:** Parallelrotation          **Abb. 4a:** Senkrechtrotation

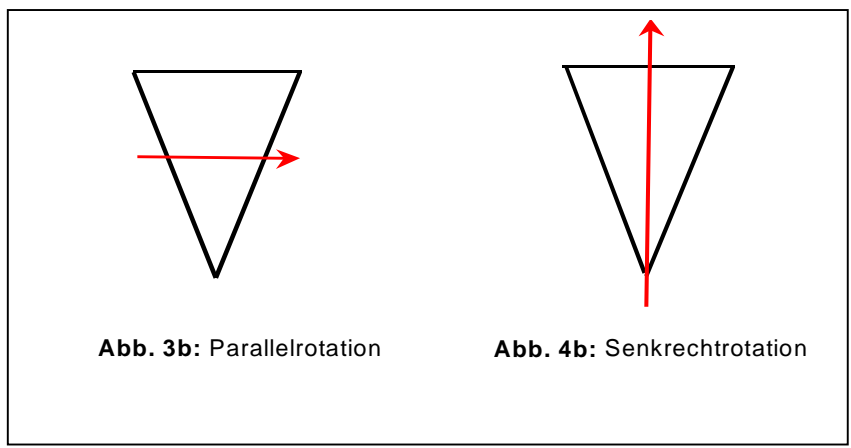

**Abb. 3b:** Parallelrotation          **Abb. 4b:** Senkrechtrotation

Für die Parallelrotation wurde die Quantenzahl $K$ eingeführt, auch Projektionsquantenzahl genannt. Für den Gesamtdrehimpuls benutzt man die Gesamtdrehimpulsquantenzahl $J$, s. Abb. 4. (Der Gesamtdrehimpuls ist häufig schräg zur Symmetrieachse des Moleküls gerichtet.) Für die Senkrechtrotation ist $(J - K)$ die zugehörige Quantenzahlkombination. ($K$ heißt Projektionsquantenzahl, weil sie zur Projektion des Gesamtdrehimpulses auf die Figurenachse gehört.) Parallel- und Senkrechtrotation können gleichzeitig stattfinden. Die kegelförmige Bewegung der Figurenachse wird übrigens Präzession genannt.

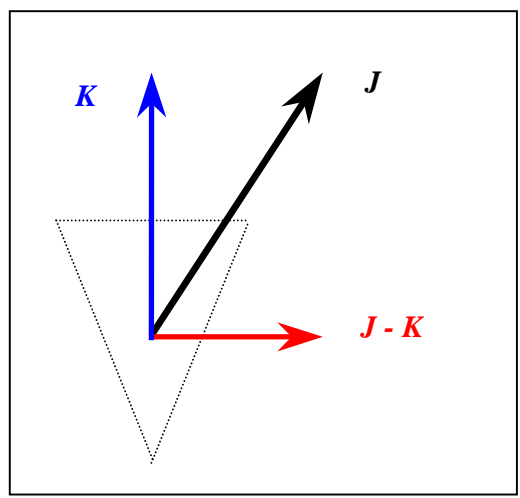

**Abb. 4**:
Die drei Quantenzahlen für die Rotation eines symmetrischen Kreisels: *J* steht für den Gesamtdrehimpuls (in beliebiger Lage zum Molekül), *K* für die Parallelrotation (Projektion des zugehörigen Vektors von *J* auf die Figurenachse) und (*J* – *K*) für die Senkrecht-rotation.

### 3.2 Aussehen eines Rotationsspektrums

Rotationsspektren von zweiatomigen Molekülen sind besonders einfach. Das liegt daran, daß diese Moleküle angeblich nur Senkrechtrotationen und keine Parallelrotation um die Stäbchenachse (Kernverbindungsachse) ausführen können, s. auch Kapitel 5.6. Deshalb wird in der folgenden Abbildung 5 als Beispiel für eine Senkrechtrotation das Rotationsspektrum von gasförmigem Chlorwasser-stoff, HCl, gezeigt. Die Auflösung ist günstiger-weise so schlecht, daß die Linien der beiden Isotopomere mit $^{35}$Cl und $^{37}$Cl nicht aufgelöst sind. Eine grobe Schnellerklärung vorweg zum Rotationsspektrum: Mit wachsender Wellenzahl lädt das Molekül je eine Stufe an Senkrecht-Rotation auf.

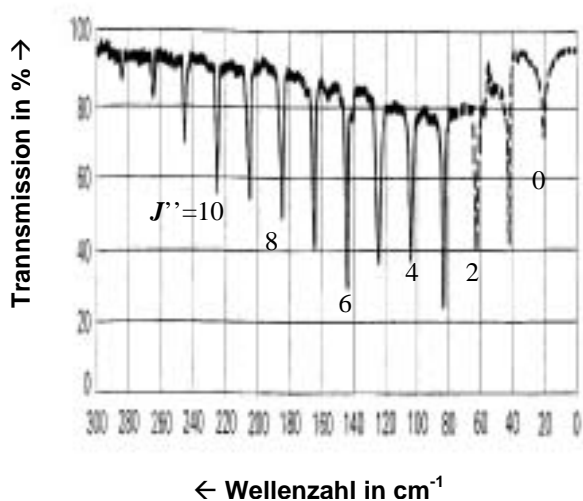

**Abb. 5**: Reines Rotations-Spektrum von gasförmigem HCl (nach DEMUTH/KOBER [3]). Die gestrichelten Linien mit der Rotationsquantenzahl im Grund-zustand *J''* = 0, 1 und 2 wurden simu-liert.

## 4. Rotations-Schwingungs-Spektroskopie

Außer der Rotation können die Moleküle auch Schwingungen ausführen. Eine Schwingung setzt sich zusammen aus zwei entgegengesetzten Rotationen. Diese beiden gegenläufigen Rotationen für ein Schwingungsquant sind in der Rotations-Schwingungs-Spektroskopie entsprechend energie-reich. Die Schwingungen erfolgen bei der Rota-tions-Schwingungs-Spektroskopie entweder senk-recht zur Figurenachse (Senkrechtschwingung oder **Senkrechtbande,** s. Abb. 6a) beziehungsweise parallel (genau genommen in) der Figurenachse (Parallelschwingung oder **Parallelbande**, s. Abb. 6b). Das sind übrigens durchaus gängige Bezeich-nungen (s. z. B. [9]). Ich zitiere übrigens bewusst das Buch von STUART, weil dieser Autor versucht hat, für einige spektroskopische Vorgänge eine klassische Erklärung zu bringen. Bei schrägem Einfall des Photons auf das Teilchen kann sowohl die Senkrecht- als auch die Parallelbande angeregt werden, sofern die Energie ausreicht. Ein Beispiel für eine Parallelbande ist in dem Buch von

DEMUTH und KOBER [3] mathematisch für Fortgeschrittene auf Schulniveau abgehandelt (aber eher wohl nur für männliche Schüler).

Die Schwingungsquantenzahl wird $v$ genannt. Bei der Absorptionsspektroskopie ist $\Delta v = 1$ und bei der Emissionsspektroskopie $\Delta v = -1$. Diese Auswahlregel wird noch einmal in Kapitel 4.2 wiederholt.

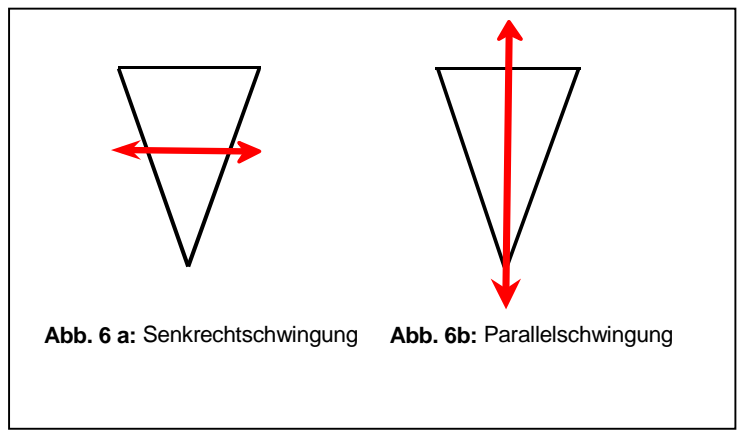

**Abb. 6 a:** Senkrechtschwingung     **Abb. 6b:** Parallelschwingung

## 4.1 Termsymbolik der Rotations-Schwingungs-Spektroskopie

Die wichtigsten Quantenzahlen der Rotations-Schwingungs-Spektroskopie haben wir kennengelernt. Sie reichen bereits aus, um eine Linie im Spektrum, einen Peak, zu kennzeichnen (zu indizieren). International üblich ist die folgende Termsymbolik:

$$^{\Delta K}\Delta J_{K''}(J'')$$

$\Delta$ bedeutet Endzustand minus Grundzustand. Für die Absorptionsspektroskopie bedeutet das: angeregter minus Ausgangszustand. $K''$ und $J''$ sind die Quantenzahlen im Ausgangszustand, normalerweise ohne Schwingungsanregung. $\Delta K$ und $\Delta J$ werden durch Großbuchstaben ersetzt. (Bitte keinen Schreck kriegen! Die Termsymbolik erscheint etwas lästig. Aber für die Forscher, die häufig mit Rotations-Schwingungs-Spektroskopie zu tun haben, bedeutet diese Symbolik durchaus eine Erleichterung.)

$\Delta K = -1$ oder $\Delta J = -1$:     P
$\Delta K = 0$ oder $\Delta J = 0$:     Q
$\Delta K = 1$ oder $\Delta J = 1$:     R

Für Parallelbanden wird das Symbol Q für $\Delta K = 0$ meist weggelassen.

## 4.2 Auswahlregeln

Hier der Vollständigkeit halber ein paar Worte zu den Auswahlregeln. Für **Rotationsspektren** gilt

$\Delta J = \pm 1$,

wobei $\Delta J = +1$ (also die Absorption) der wohl am häufigsten beobachtete Fall ist.

Bei der **Rotations-Schwingungs-Spektroskopie** kommt wegen der Schwingung die Schwingungsquantenzahl $v$ ins Spiel:

$\Delta v = \pm 1$.

Außerdem spielt die Quantenzahl $K$ für die Rotation um die Figurenachse häufig eine Rolle, zwar nicht für **Parallelbanden**, denn für diese gilt

$\Delta K = 0$,

statt dessen gehorchen **Senkrechtbanden** der Auswahlregel

$\Delta K = \pm 1$

und $\Delta J = \pm 1$ sowie zusätzlich auch noch $\Delta J = 0$.

Zweiatomige Moleküle liefern nur Parallelbanden und mehratomige Moleküle zusätzlich zu den Parallelbanden auch noch Senkrechtbanden.

Senkrechtbanden sind wegen der Linienvielfalt etwas komplizierter und nicht geeignet für eine Behandlung in der Schule. Ich werde dieses Thema deshalb nur am Rande (wegen der Vollständigkeit) streifen.

| Rotations-Spektroskopie $\Delta v = 0$; $\Delta J = \pm 1$ | |
| --- | --- |
| Rotations-Schwingungs-Spektroskopie $\Delta v = \pm 1$ | |
| Parallelbande $\Delta K = 0$; $\Delta J = (0),\pm 1$ | Senkrechtbande $\Delta K = \pm 1$; $\Delta J = 0,\pm 1$ |

**Tab. 1**: Einige wichtige Auswahlregeln für symmetrische Kreisel

Über die Auswahlregel $\Delta J = 0$ für eine Parallelbande s. kurz Kapitel 5 und 5.4.

## 4.3 Aussehen von Rotations-Schwingungs-Spektren

In den folgenden beiden Abbildungen wird je ein experimentelles Beispiel für eine Parallelbande von HCl (in Abb. 7) und eine Senkrechtbande von $CH_3Cl$ (in Abb. 8) gezeigt. Der R-Zweig im Rotations-Schwingungs-Spektrum von HCl in Abb. 7 ähnelt vom Aussehen her, wenn man von der Auflösung abstrahiert, dem reinen Rotationsspektrum von HCl in Abb. 5. Das gilt allgemein für die R-Zweige innerhalb der Parallelbanden von zweiatomigen Molekülen bis hin zu symmetrischen Kreiselmolekülen. Im Gegensatz zum Rotationsspektrum der Abb. 5 ist in Abb. 7 die Auflösung des Rotations-Schwingungs-Spektrums so gut, daß die Peaks der beiden unterschiedlich schweren Isotopomere mit $^{35}Cl$ und $^{37}Cl$ aufgelöst sind. Vergleiche zusätzlich auch noch die schematische Abbildung 9, und zwar die erste Reihe mit $K = 0$, die typisch ist für das Rotations-Schwingungs-Spektrum eines zweiatomigen Moleküls.

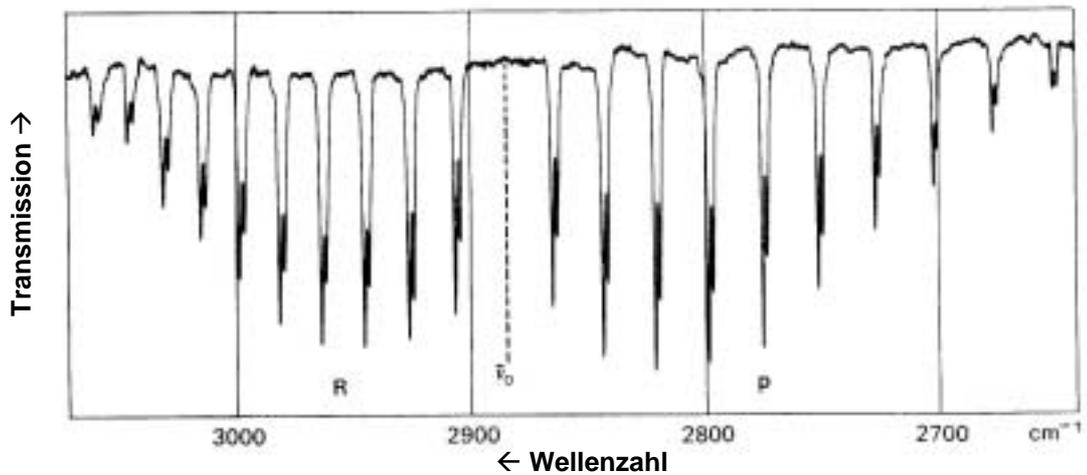

**Abb. 7**: Beispiel für eine Parallelbande ist das Rotations-Schwingungs-Spektrum von HCl (aus DEMUTH/KOBER [3])

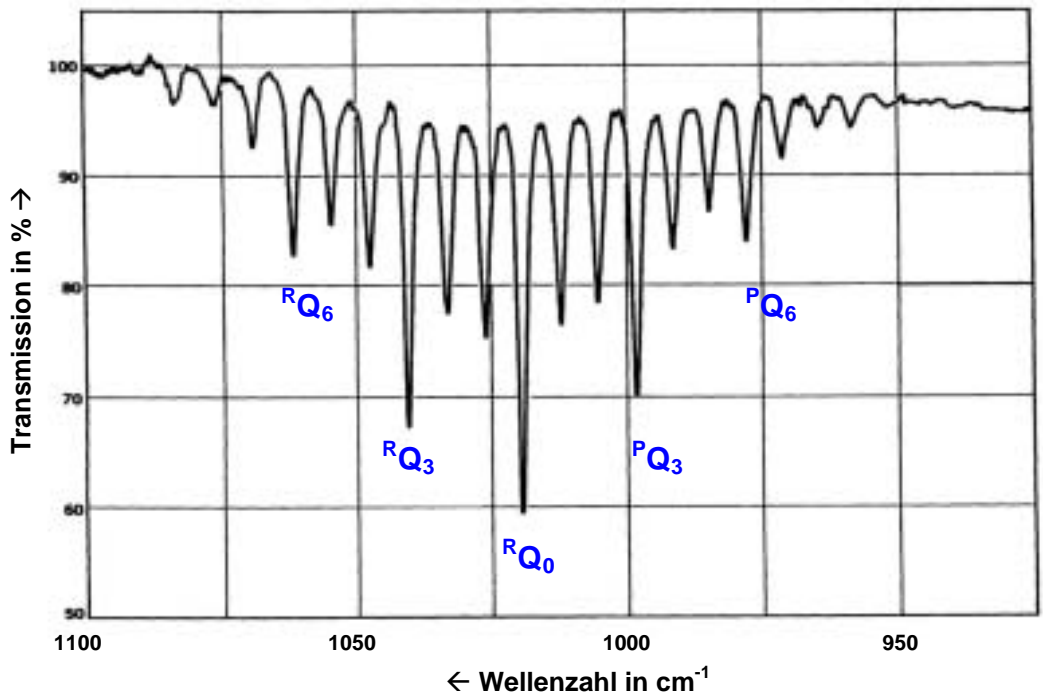

**Abb. 8**: Beispiel für eine Senkrechtbande ist das Rotations-Schwingungs-Spektrum von $CH_3Cl$ bei schwacher Auflösung (aus BARROW [1])

Wegen der schwachen Auflösung bei der Senkrechtbande von $CH_3Cl$ in Abb. 8 sind nur die $^RQ$- und $^PQ$-Zweige zu erkennen. Das hat folgenden Grund: Die zu einem $K$-Wert gehörenden Linien mit unterschiedlichem $J$ liegen fast an der gleichen Stelle auf der Wellenzahl-Achse und addieren ihre Intensität. Die übrigen Linien gehen im Rauschen unter.

## 5. Deutung von spektroskopischen Übergängen

Woran erkennt man denn die Schwingung? Die Schwingung sieht man nicht direkt im Spektrum. Sie lässt sich nur aus der Lage der Peaks erahnen. In Abb. 7 ist der Bandenursprung der Schwingung als $\nu_0$ markiert. Rotations-Übergänge dagegen sieht man. Sie sind sehr ähnlich den Rotations-, den Rotations-Schwingungs- und den Elektronen-Schwingungs-Rotations-Spektren (sofern letztere so gut aufgelöst sind).

An dieser Stelle sei nun verraten, was bei den einzelnen Zweigen einer Parallelbande ($^QQ$-, $^QR$-, $^QP$-Zweig) und einer Senkrechtbande ($^RR$-, $^PP$-, $^RQ$-, $^PQ$-, $^RP$-, $^PR$-Zweig) im einzelnen für Rotationsänderungen auftreten, weil dies in keinem Lehrbuch steht, s. Tab. 2 und Tab. 3.

Ob bei einer Parallelbande ein $^QQ$-Zweig sichtbar ist, darüber entscheiden die Kernspins in den Molekülen und damit die Kernspinstatistik. Aber hierauf soll nicht eingegangen werden.

**Tab. 2**: Was passiert bei der Parallelbande außer der Schwingung?

| $^QQ$ (paralleler und senkrechter) Rotationszustand bleibt konstant | |
|---|---|
| $^QR$ Senkrechtrotation aufladen | $^QP$ Senkrechtrotation abladen |

**Tab. 3**: Was passiert bei einer Senkrechtbande außer der Schwingung?

| $^RR$ Parallelrotation aufladen | $^PP$ Parallelrotation abladen |
|---|---|
| $^RQ$ Senkrecht- in Parallelrotation umwandeln | $^PQ$ Parallel- in Senkrechtrotation umwandeln |
| $^RP$ Senkrecht- in Parallelrotation umwandeln und Senkrechtrotation abgeben | $^PR$ Parallel- in Senkrechtrotation umwandeln und Senkrechtrotation aufladen |

### 5.1 Zur Lage der ersten Linien

Im folgenden wird zur Vereinfachung gesetzt $K = K''$. Es werden die Spektren von symmetrischen Kreiselmolekülen betrachtet. Wo liegen nun die ersten beobachtbaren Linien? Die ersten Linien einer Parallelbande lauten ab $K = 0$:

$$^QR_K(K)$$

$$^QQ_K(K) \; [^QQ_K(K) \text{ erst ab } K=1]$$

$$^QP_K(K+1).$$

(Achtung, die Ausdrücke in den runden Klammern geben für einen vorgegebenen Wert von $K$ den kleinsten erlaubten Wert für $J$ an.)

Als erste Linien einer Senkrechtbande im P-Zweig sind erlaubt ab $K=1$:

$$^PR_K(K)$$

$$^PQ_K(K)$$

$$^PP_K(K),$$

und im R-Zweig schon ab $K=0$:

$$^RR_K(K)$$

$$^RQ_K(K+1)$$

$$^RP_K(K+2).$$

Hier nur kurz zum erläuternden Verständnis ein Beispiel: Die ersten Linien einer Parallelbande

lauten für $K = 0$: ${}^{Q}R_0(0)$ und ${}^{Q}P_0(1)$, s. Abb. 9 in der ersten Reihe. Auch wenn keine Senkrechtrotation vorhanden ist wie bei $J = 0$, kann eine solche aufgeladen werden: Das entspricht dem Übergang ${}^{Q}R_0(0)$. Eine Senkrechtrotation kann erst abgegeben werden, wenn eine solche existiert, also ab $J = 1$, dieses ist der Übergang ${}^{Q}P_0(1)$.

Für Eingeweihte noch eine analoge Ergänzung aus der Raman-Spektroskopie: Für $K = 0$ gibt es am absoluten Nullpunkt keine Emissionslinie (Anti-Stokes-Linie), sondern nur eine Absorptionslinie (Stokes-Linie).

Eine Parallelbande kann z. B. bei guter Auflösung schon sehr linienreich sein, s. andeutungsweise Abb. 9, letzte Zeile. Man kann in diese Linienvielfalt eine gewisse Ordnung und Vereinfachung hineinbringen, wenn man Teile des Spektrums nach $K$-Werten im Grundzustand zusammenfasst. Diese Teile werden als Subbanden bezeichnet und bestehen mindestens aus einem P- und einem R-Zweig. Solch ein Spektrum lässt sich in Subbanden für verschiedene $K$-Werte eines mehratomigen Moleküls unterteilen. Beim zweiatomigen Molekül lässt sich nur die einzige Subbande mit $K=0$ beobachten.

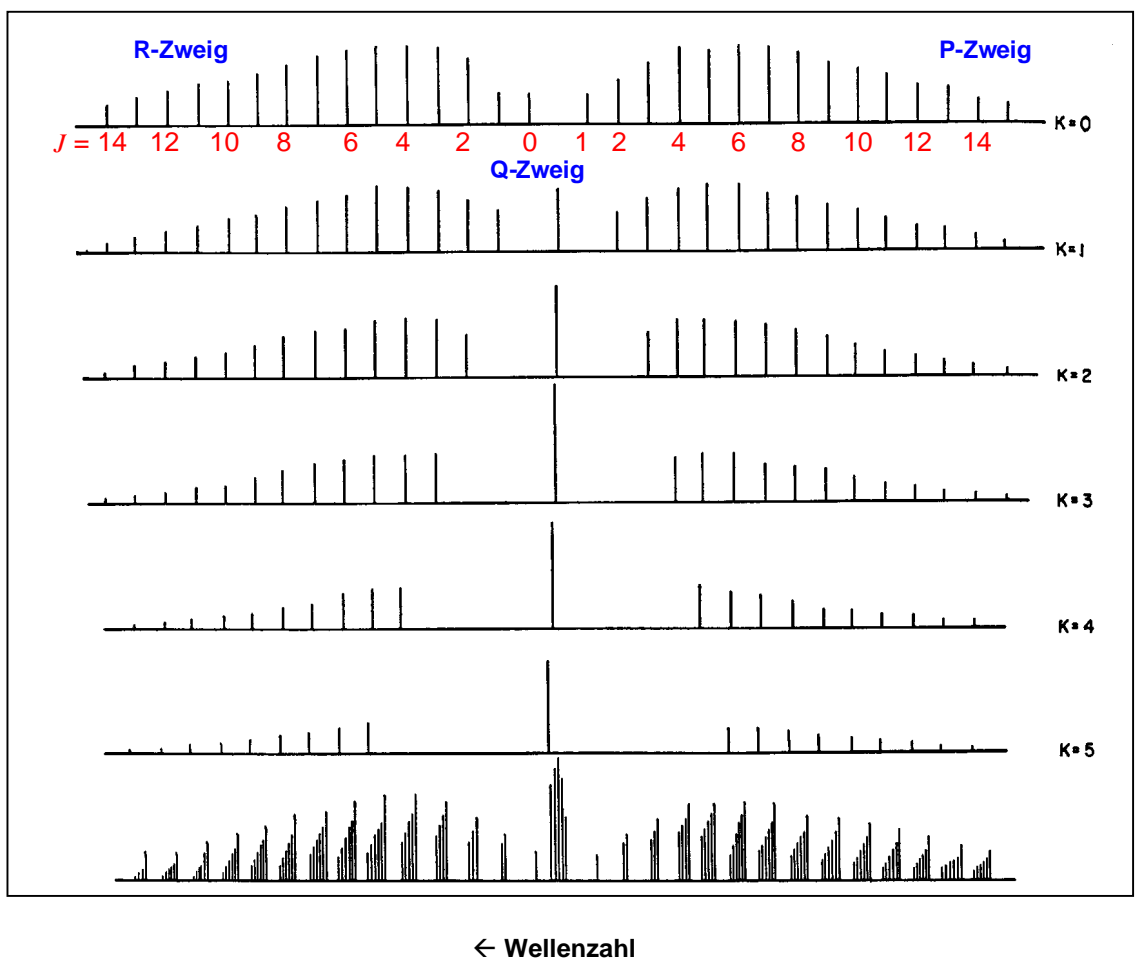

← **Wellenzahl**

**Abb. 9**: Die Subbanden einer Parallelschwingung für $K=0$ bis $K=5$ mit $J=0$ bis maximal 15. Im untersten Spektrum sind alle genannten Subbanden aufaddiert (nach CONDON [2]). Ein Q-Zweig ist nur für bestimmte Kernspin-Konstellationen bei einem symmetrischen Kreisel zu beobachten.

## 5.2 Zur möglichen Deutung von Schwingungen

Eine Schwingung ist prinzipiell denkbar durch eine starke Rotation, bei der sich außerdem eine gleichstarke Rotation in entgegengesetzter Richtung ereignet. Das wurde bereits in Kapitel 4 erwähnt. Nicht eingehen möchte ich auf die durchaus interessanten Fälle, bei denen sich Spins oder schwache Rotationen aufheben und eine schwachen Schwingungszustand ergeben. Die folgenden beiden Sätze sind nur für speziell eingeweihte Personen gedacht: Für die IR-Spektroskopie bedeutet das, daß man für die nächst höhere Schwingungsanregung geringfügig weniger Energie benötigt (Anharmonizität). In der Elementarteilchenphysik liefern Schwingungsanregungen Extra-Beiträge zum Formfaktor eines Kerns [10].

## 5.3 Zur Vielzahl der Moleküle

Die Suche nach der reellen Gesamtwellenfunktion ist ein schwieriges Puzzle. Ein Problem stellt das Vorhandensein von sehr vielen Molekülen dar und ihren Aufeinanderwirken innerhalb eines großen Beobachtungs-Zeitraums. Ein einzelnes Molekül liefert bei Photonenbeschuss nur einen kleinen Peak. Ich wiederhole noch einmal: Wenn die Wellenfunktion eines einzelnen Moleküls reell ist, muss natürlich die Gesamtwellenfunktion des Teilchenensembles auch reell sein.

Wegen des großen Beobachtungszeitraums sehen wir quasi „gleichzeitig" die Gesamtleistung von vielen Molekülen und deren Isotopomeren, also ganz viele Peaks. Aber wir sehen nicht, welche Übergänge erlaubt sind und wie die Peaks entstanden sind. In einem Punkt können wir etwas sicher sein (es gilt die „Verhältnismäßigkeit der Mittel", sprich der Energie): Normalerweise wird sich ein Isomer kaum in ein anders verwandeln („transmutieren"), es sei denn in der Spektroskopie der Hochenergiephysik. Achtung, praktisch die gesamte Elementarteilchenphysik ist nichts anderes als Hochenergie-Spektroskopie.

## 5.4 Zum Kernspin

Der Kernspin entscheidet über die Existenz eines Q-Zweiges bei einer Parallelbande sowie über das Intensitätsverhältnis der Linien in einer Subbande einer Parallel- oder einer Senkrechtschwingung. Deshalb sind z. B. bei Molekülen mit einer dreizähligen Drehachse wie $CH_3Cl$ die Linien mit $K$-Werten, die durch 3 teilbar sind, doppelt so stark wie die übrigen Linien (s. Abb. 8). Allgemein verständliche Hinweise zum fast gleichen Thema findet man in den Lehrbüchern für Chemiker in den Kapiteln zur Kernresonanzspektroskopie (NMR-Spektroskopie) an Wasserstoffkernen.

Bei der mathematischen Modellierung der Rotations-Schwingungs-Spektroskopie spielt heutzutage die Kernspinquantenzahl eine kleine Rolle, da die absolute Auslösung der Spektren noch nicht so fein ist wie bei der NMR-Spektroskopie.

## 5.5 Zur Gravitation

Wer es mag, kann die Linien eines Spektrums im Kapitel Quantengravitation abheften und ist damit über das Wesen physikalischer Vorgänge für´s erste ganz gut informiert.

## 5.6 Die „heilige Kuh" der Auswahlregeln

Die Gültigkeit der Auswahlregeln ist relativ und zwar stark abhängig von den Versuchsbedingungen.

Das Kapitel Auswahlregeln ist leider in den Lehrbüchern qualitativ überhaupt nicht verstanden und deshalb nur mühselig nachvollziehbar gemäß den Aussagen der Quantentheorie. Der Versuch einer Erklärung durch HERRMANN [4] ist lobenswert, aber leider doch zu einfach gefasst. Ich selbst kann aber nur ein paar Ergänzungen anbringen:

Es sieht ganz danach aus, daß die untersuchten spektroskopischen Systeme immer dann besonders stark eine Auswahlregel mißbachten, wenn sie weit vom thermischen Gleichgewicht entfernt sind.

So erlauben Versuchsbedingungen der Astrophysik verbotene Linien durch die extrem schwache Teilchendichte (keine Stöße zwischen den Teilchen) und die riesigen Durchstrahlungsweg zu beobachten. Unter irdischen Laborbedingungen machen hochsymmetrische Teilchen, Elektronen oder Moleküle elastische Stöße untereinander, so dass am Detektor keine Änderungen registriert werden. Unsymmetrische gleichartige Moleküle mögen unter sich noch teilweise elastische Stöße ausführen können, aber an den Wänden werden die angeregten Moleküle ihre Überschussenergie nicht los wegen der unpassenden Figur der Wandmoleküle, so dass von diesen unsymmetrischen Molekülen ein Peak registriert werden kann.

Ferner kann eine hohe Intensität während kurzer Messzeit dafür sorgen, dass verbotene Linien im Spektrum auftauchen.

Auch das Hinzumischen von Fremdteilchen kann Wunder bewirken (Darauf beruht z. B. die Funktionsweise des Helium-Neon-Lasers).

Wenn im Beobachtungszeitraum genauso viele Moleküle an- wie abgeregt sind, so wird man überhaupt keinen Peak im Spektrum sehen, das ist sozusagen eine direkte Folge der Auswahlregeln. Das ist z. B. der Fall bei gleichartigen symmetrischen Teilchen (z. B. Rotationen von Elektronen, Atomen sowie homonuklearen zweiatomigen Molekülen).

Zur Auswahlregel $|\Delta J| > 1$ bei Rotationsänderungen: Auf der DPG-Tagung in Berlin 1997 [6] habe ich auf die Beobachtbarkeit von verbotenen Linien aus neuerer Zeit bei Silylhalogeniden mit $\Delta K = \pm 3$ hingewiesen. Inzwischen wurde auch schon die Auswahlregel $\Delta K = \pm 6$ beobachtet [5]. Trotz dieser neu entdeckten Auswahlregeln wird die Spektroskopie z. Zt. noch nicht revolutioniert.

Hier möchte ich noch eine diesbezügliche Vermutung äußern: Kernspin-Übergänge und Rotations-Übergänge benötigen vergleichsweise wenig Energie im Gegensatz zur Schwingungsanregung. Deshalb sind Kernspin und Rotationen bis zu viel höheren Quantenzahlen angeregt, als man bisher glaubte. Die Rotations-Schwingungs-Photonen werden meist durch fast elastische Stöße von Teilchen zu Teilchen übertragen. Diese Art der Stöße ist deshalb im Spektrum beobachtbar, weil das mit dem Photon beladene Teilchen mit höherer Wahrscheinlichkeit auf ein anderes Teilchen stößt, das sich eher in einem anderen Kernspin- oder/und Rotationszustand befindet. Ich glaube, was ich eben so lapidar dahin geschrieben habe, scheint ganz wichtig für die Beobachtung von Spektren zu sein.

Kleinere Auflösungen, schnellere Computer werden in Zukunft immer noch weitere neue Auswahlregeln hervorbringen und für frappierende Überraschungen sorgen. Je besser spektroskopische Übergänge gedeutet werden können und das entsprechende Bewußtsein dafür vorhanden ist, um so mehr findet die **Determiniertheit im mikrokosmischen Bereich** wieder Einzug in die Physik. Bis dahin ist aber noch ein weiter beschwerlicher Weg zu beschreiten.

## 6. Fazit

Wie können wir uns das Zustandekommen einer Absorptionslinie bei der IR-Spektroskopie prinzipiell erklären? Unumstritten ist die Tatsache, dass ein Photon auf das Molekül einen Drehimpuls abwälzt. Gleichzeitig wird bei der Rotations-Schwingungs-Spektroskopie auch noch ein Schwingungsimpuls mit abgeladen. Um im feldfreien Raum eine Spektrallinie beobachten zu können, benötigt man Teile in einem Molekül, die drehbar sind, und die aus dem Photon Drehimpuls für den eigenen Bedarf zur Änderung von Spin, Rotation, Kernspin aufnehmen können. Es führt zu einer Intensitätserhöhung, wenn Übergänge bei fast der gleichen Wellenzahl auftreten, und das kann der Grund dafür sein, daß Übergänge überhaupt erst sichtbar werden.

## 7. Literatur

[1] BARROW, G. M.: *Introduction of Molecular Spectroscopy*. Tokio: McGraw-Hill 1962

[2] CONDON, E. U. (Herausgeber): *Handbook of Physics*, 2. Aufl. New York: McGraw Hill 1967

[3] DEMUTH, R.; KOBER, F.: *Grundlagen der Spektroskopie*. Frankfurt: Diesterweg-Salle 1977

[4] HERRMANN, F.: Auswahlregeln, verbotene Übergänge. In: *Physik in der Schule* 38, (2000), S. 74

[5] RULAND, H.: Dissertation. *Rotations-Schwingungsanalyse symmetrischer und asymmetrischer Kreiselmoleküle nahe dem Kugelkreisel*. Wuppertal 2000
http://www.bib.uni-wuppertal.de/elpub/fb09/diss2000/ruland/d090008.pdf
Stand: Juli 2003

[6] SCHULZ, P.: Kurioses zum Photon. In: Tagungsband *Didaktik der Physik, Vorträge*. Berlin: Deutsche Physikalische Gesellschaft, 1997

[7] SCHULZ, P.: Plausible Definition von Masse, Ladung und Spin, DPG-Vorträge Didaktik der Physik und Teilchenphysik, März 2000 in Dresden. In: *CD zur Frühjahrstagung Didaktik der Physik in der Deutschen Physikalischen Gesellschaft*, Dresden 2000

[8] SCHULZ, P.: Plausible Erklärungshinweise gegen die Überlichtgeschwindigkeit, DPG-Vortrag Didaktik der Physik, März 2001 in Bremen. In: *CD zur Frühjahrstagung Didaktik der Physik in der Deutschen Physikalischen Gesellschaft*, Bremen 2001

[9] STUART, H. A.: *Molekülstruktur*, 3. Auflage. Berlin: Springer 1967

[10] GREULICH, W. (Herausgeber): *Lexikon der Physik*: Heidelberg: Spektrum Akademischer Verlag, 1998